\documentclass[aps,prl,reprint,superscriptaddress]{revtex4-1}

\usepackage{graphicx}
\usepackage{amssymb, amsmath}
\usepackage[usenames]{color}
\usepackage{dcolumn}
\usepackage{bm}
\usepackage{ textcomp }
\usepackage{url}

\begin{document}

\title{Black Hole Horizon in a Type-III Dirac Semimetal Zn$_2$In$_2$S$_5$}

\author{Huaqing Huang}
\affiliation{Department of Materials Science and Engineering, University of Utah, Salt Lake City, Utah 84112, USA}

\author{Kyung-Hwan Jin}
\affiliation{Department of Materials Science and Engineering, University of Utah, Salt Lake City, Utah 84112, USA}

\author{Feng Liu\footnote{Corresponding author: fliu@eng.utah.edu}}
\affiliation{Department of Materials Science and Engineering, University of Utah, Salt Lake City, Utah 84112, USA}
\affiliation{Collaborative Innovation Center of Quantum Matter, Beijing 100084, China}

\date{\today}

\begin{abstract}
Recently, realizing new fermions, such as type-I and type-II Dirac/Weyl fermions in condensed matter systems, has attracted considerable attention. Here we show that the transition state from type-I to type-II Dirac fermions can  be viewed as a type-III Dirac fermion, which exhibits unique characteristics, including a Dirac-line Fermi surface with nontrivial topological invariant and critical chiral anomaly effect, distinct from previously known Dirac semimetals. Most importantly, we discover Zn$_2$In$_2$S$_5$ is a type-III Dirac semimetal material, characterized with a pair of Dirac points in the bulk and Fermi arcs on the surface. We further propose a solid-state realization of the black-hole-horizon analogue in inhomogeneous Zn$_2$In$_2$In$_5$ to simulate black hole evaporation with high Hawking temperature. We envision that our findings will stimulate researchers to study novel physics of type-III Dirac fermions, as well as astronomical problems in a condensed matter analogue.
\end{abstract}



\maketitle

The conception of topology \cite{PhysRevLett.49.405,PhysRevLett.50.1153,kosterlitz1972long,*kosterlitz1973ordering} has been known in condensed matter physics since 1980¡¯s. However, it is until recently that we have witnessed an exponential growth in the field of topological phases of mater in the last decade; thanks to the introduction of concept of topological insulators (TIs) \cite{PhysRevLett.95.146802,*PhysRevLett.95.226801} and their proposition \cite{bernevig2006quantum} and confirmation \cite{konig2007quantum} in real material systems. This effective route to discovering TIs from theoretical conception to computational material proposition and to experimental confirmation has been followed by the discovery of other topological materials, such as topological crystalline insulator \cite{PhysRevLett.106.106802,hsieh2012topological,tanaka2012experimental,*dziawa2012topological,*okada2013observation}, Dirac semimetal \cite{Na3Bi,Na3Bi_exp1,*Na3Bi_exp2, Cd3As2,Cd3As2_exp1,*Cd3As2_exp2}, and most recently Wyel semimetal \cite{burkov2011topological,PhysRevB.83.205101,TaAsPRX,*TaAsHLin, xu2015discovery,*yang2015weyl,*lv2015observation}. Depending on the geometry of Dirac cone, there are type-I and type-II Dirac/Weyl semimetals \cite{type2Weyl,shuyunMoTe2,huanghqPtSe2,PtTe2,chang2016type}. Interestingly, it has been shown that the interface or the ``transition state'' from type-I to type-II has distinctly different topological properties \cite{volovik2016black,volovik2016lifshitz,volovik2017exotic}, which we will call it a ``type-III'' semimetal. Both type-I and type-II Dirac/Weyl semimetals have been experimentally confirmed in real materials \cite{Na3Bi_exp1,*Na3Bi_exp2,Cd3As2_exp1,*Cd3As2_exp2,xu2015discovery,*yang2015weyl,*lv2015observation,shuyunMoTe2,PtTe2}; however, so far type-III Dirac/Weyl semimetal remains a theoretical conception. In this Letter, we will fill this outstanding gap by proposing a type-III semimetallic phase in a real material Zn$_2$In$_2$S$_5$.

Topological semimetals host interesting new types of fermions as low-energy quasiparticles.
They not only exhibit novel physical properties such as distinctive topological surface states \cite{Na3Bi, Na3Bi_exp1,*Na3Bi_exp2, Cd3As2, Cd3As2_exp1,*Cd3As2_exp2, burkov2011topological,PhysRevB.83.205101,TaAsPRX,*TaAsHLin,xu2015discovery,*yang2015weyl,*lv2015observation, type2Weyl},
large linear magnetoresistance \cite{PhysRevB.58.2788, shekhar2015extremely, liang2015ultrahigh} and chiral anomaly \cite{nielsen1983adler,wang2016gate, xiong2015evidence, li2015giant,*li2016negative, zhang2016signatures,*PhysRevX.5.031023,huanghqSn}, but also offer a versatile platform for simulating relativistic particles of high-energy physics as well as ``new particles'' that have no counterparts in high-energy physics. The type-III Dirac semimetal has been theoretically proposed for realizing a solid-state analogue of block hole horizon \cite{volovik2016black,volovik2016lifshitz,volovik2017exotic}. To this end, we will again fill the gap by devising a material platform, an inhomogeneous Zn$_2$In$_2$S$_5$, to simulate Hawking radiation at the black-hole horizon. Especially we suggest a high Hawking temperature associated with the analogous black-hole horizon in Zn$_2$In$_2$S$_5$ to ease the experimental observation, in contrast to the low Hawking temperature in previously proposed black-hole-horizon analogues, such as sound wave in flowing medium \cite{PhysRevLett.46.1351, *PhysRevD.51.2827, *visser1998acoustic}, superfluid helium \cite{PhysRevD.58.064021,*volovik2001superfluid,volovik2003universe} and Bose-Einstein condensates \cite{PhysRevLett.85.4643,steinhauer2014observation,*steinhauer2016observation}.

We will first highlight the key features of the type-III Dirac semimetals, including their unique characteristics of Dirac-line Fermi surface with nontrivial topological invariant and critical chiral magnetic effect, in distinction from those of type-I and type-II Dirac semimetals. Then we will present evidence that Zn$_2$In$_2$S$_5$ is the first candidate material for realizing the type-III Dirac fermions. Based on effective Hamiltonian analysis and first-principles calculations, we show novel properties of Zn$_2$In$_2$S$_5$ including critical chiral magnetoresponse and Fermi arcs. Finally we will describe a solid-state realization of the black-hole-horizon analogue in inhomogeneous
Zn$_2$In$_2$S$_5$, to simulate black hole evaporation with a relatively high Hawking temperature. Our work not only enriches the fundamental physics of topological semimetals with new types of fermions but also provides an alternative route to studying black hole horizon, in real materials systems.

\begin{table*}
\caption{\label{TabI}  Comparison among three types of Dirac semimetals.}
\begin{tabular}{l|lll}
  \hline
  \hline
   &  Type-I &  Type-II &  Type-III \\
  \hline
   Dispersion &  Dirac cone &  Overtilted Dirac cone   &  Critical Dirac cone\\
   Fermi surface  &  Point-like &  Electron \& hole pockets &  Dirac line (with $N_2=1$)\\
   DOS($E_F$)            &  Vanishing &   Parabolic peak       &  Finite \\
   Surface Fermi Arc &   \checkmark  &       \checkmark                    &   \checkmark \\ 
   Chiral anomaly  &   Along all direction &  Anisotropic, inside a cone region\cite{PhysRevLett.117.077202,*PhysRevLett.117.086401,*PhysRevLett.117.086402}  &  Except for the critical plane [41]\\
   Black hole analogue &  Outside &  Inside &  Horizon\\
   Typical materials &  Na$_3$Bi\cite{Na3Bi}, Cd$_3$As$_2$\cite{Cd3As2} &  PtTe$_2$\cite{huanghqPtSe2,PtTe2}, VAl$_3$\cite{chang2016type} &  Zn$_2$In$_2$S$_5$ (this work) \\
  \hline
  \hline
\end{tabular}
\end{table*}

\begin{figure}
\includegraphics[width =1\columnwidth]{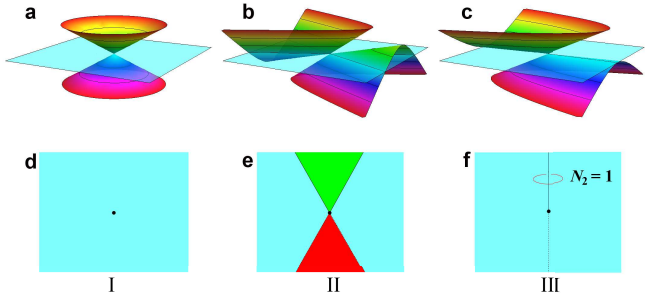}%
\caption{\label{fig1_type3} (a) Type-I Dirac point with a point-like Fermi surface. (b) Type-II Dirac point is the contact point between electron and hole pocket. (c) Type-III Dirac point appears as the touching point between Dirac lines. The Dirac line is described by the topological invariant $N_2\!=\!1$. The light blue semitransparent plane corresponds to the position of the Fermi level, and the black solid/dashed lines mark the boundary of hole/electron pockets. (d)-(f) Fermi surfaces of three types of Dirac semimetals.
}
\end{figure}

Topological Dirac and Weyl semimetals are characterized with fourfold and twofold linear band crossings at the Fermi level (the so-called Dirac and Weyl points), respectively. They can be further classified into two types by fermiology. One is the type-I Dirac/Weyl semimetals which have a typical conical dispersion and point-like Fermi surface [Fig.~\ref{fig1_type3}(a)] \cite{Na3Bi,Na3Bi_exp1,*Na3Bi_exp2, Cd3As2,Cd3As2_exp1,*Cd3As2_exp2, TaAsPRX,*TaAsHLin,xu2015discovery,*yang2015weyl,*lv2015observation}. The other is the type-II Dirac/Weyl semimetals which manifest in an overtilted cone-shape band structure, possessing both electron and hole pockets that contact at the type-II Dirac/Weyl point [Fig.~\ref{fig1_type3}(b)] \cite{type2Weyl,shuyunMoTe2,huanghqPtSe2,PtTe2,chang2016type}. The type-III Dirac semimetal is distinct from both type-I and type-II semimetals. As illustrated in Fig.~\ref{fig1_type3}(c), the type-III Dirac point is also a protected band crossing point, but appears at the contact of a line-like Fermi surface. Unlike Fermi surfaces of other topological semimetals, such a unique line-like Fermi surface, so-called Dirac line, is protected by a topological invariant which is an integer only for the relatively rare Dirac line.

Since a Dirac point can be viewed as the merge of a pair of Weyl points with opposite chirality, we start by considering a general $4\times4$ Hamiltonian composed of two $2\times2$ Hamiltonian describing Weyl points for simplicity,
\begin{equation}
H(\mathbf{k})=\left(\begin{array}{cc}
h(\mathbf{k})&0\\
0&h^*(-\mathbf{k})\\
\end{array}
\right)
\end{equation}
with
\begin{equation}
h(\mathbf{k})=\mathbf{v}\cdot \mathbf{k}\sigma_0+\sum_{i,j}k_iA_{ij}\sigma_j,
\end{equation}
where $\sigma_j$ are Pauli matrices and $\sigma_0$ is the identity matrix. The energy spectrum of a Weyl point is $E_{\pm}(\mathbf{k})= \sum_iv_ik_i \pm\sqrt{\sum_j\left(\sum_{i}k_iA_{ij}\right)^2}=T(\mathbf{k})\pm U(\mathbf{k})$. It is well-known that if there exists a direction for which $T>U$, the band crossing point is a type-II Dirac point, otherwise it is a type-I Dirac point. If and only if for a particular direction $\hat{k}$ in reciprocal space, $T(\hat{k})=U(\hat{k})$, but $T(\hat{k})<U(\hat{k})$ for other directions, the Dirac points are connected by a line-like Fermi surface which is the Dirac line of the type-III Dirac semimetal. It is distinctively different from the type-I case of point-like Fermi surface or the type-II case of hyperbolic Fermi surface (coexistence of electron and hole pockets). The Dirac line is protected by the combination of topology and symmetry, and can be described by a topological invariant \cite{volovik2016lifshitz},
\begin{equation}
N_2=\frac{1}{4\pi i}\mathrm{Tr}[K \oint _c dl h(\mathbf{k})^{-1}\partial_l h(\mathbf{k})],
\label{N2}
\end{equation}
where $C$ is a contour enclosing the Dirac line in momentum space. Here $K$ is a proper symmetry operator, which commutes or anticommutes with the Hamiltonian (see Supplemental Material for more details \footnote{\label{fn}See Supplemental Material at http://link.aps.org/supplemental/xxx, for more details about the computation, which include Refs.~\cite{mp,VASP,PBE,wannier90,lopez,*lopez2}.}).
The topological invariant is actually a winding number of phase around the line, and it stabilitzes the Dirac line in the sense that the integral is integer, $N_2\!=\!1$ only for type-III semimetal. The Dirac line in momentum space with nonzero winding number is an analogue of the vortex line in superfluids \cite{volovik2003universe}.


Interestingly, the type-III Dirac semimetal can also be viewed as the critical state of Lifshitz transition between type-I and type-II Dirac semimetals \cite{volovik2016lifshitz}. The Lifshitz transition was investigated recently and a solid-state realization of black-hole-horizon analogue based on inhomogeneous topological semimetals was proposed \cite{volovik2016black,volovik2016lifshitz,volovik2017exotic}.
It is hoped that topological semimetals will provide an alternative way to observe black hole horizon in condense matter systems. So far, however, no material system is known to be a type-III Dirac semimetal. Next, we will fill this gap by demonstrating Zn$_2$In$_2$S$_5$ to be the first type-III Dirac semimetal, and how to realize its black-hole-horizon analogue.

Zn$_2$In$_2$S$_5$ has a layered structure consisting of nonuple layers stacked together along the $z$-direction (see Supplemental Material \footnotemark[\value{footnote}]). Each nonuple layer consists of two In and two Zn layers which are sandwiched by S layers alternately, and every In or Zn atom lies in the center of a terahedron or octahedron of S atoms. The coupling is strong between atomic layers within the nonuple layer but much weaker between adjacent nonuple layers. Due to different stacking of these basic building blocks, two kinds of structures of Zn$_2$In$_2$S$_5$ arise, i.e., AB-stacked hexagonal structure with $P6_3mc$ symmetry and ABC-stacked rhombohedral structure with $R3m$ symmetry.

\begin{figure}
\includegraphics[width =1\columnwidth]{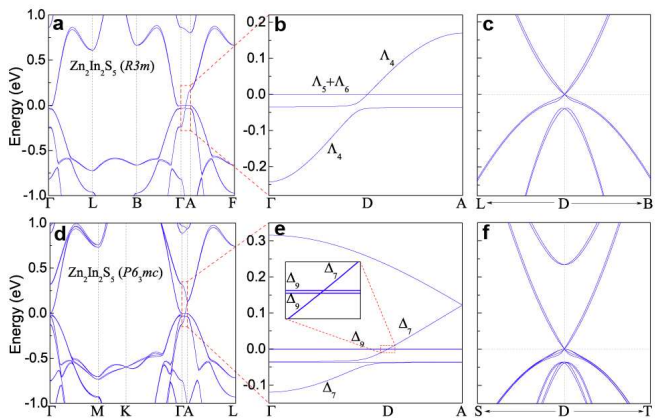}%
\caption{\label{fig2_band} Band structures of (a)-(c) Zn$_2$In$_2$S$_5$($R3m$) and (d)-(f) Zn$_2$In$_2$S$_5$($P6_3mc$). (b) and (e) The zoom-in band structures along $\Gamma$-A of Zn$_2$In$_2$S$_5$($R3m$) and Zn$_2$In$_2$S$_5$($P6_3mc$), respectively. (c) and (f) The in-plane band dispersions around the Fermi level of Zn$_2$In$_2$S$_5$($R3m$) and Zn$_2$In$_2$S$_5$($P6_3mc$), respectively.
}
\end{figure}

We first calculated the band structure of Zn$_2$In$_2$S$_5$. As shown in Fig.~\ref{fig2_band}(a) and~\ref{fig2_band}(d), there are flat bands along the $\Gamma$-A direction near the Fermi level. Meanwhile, another band disperses upward crossing the flat bands in between $\Gamma$ and A.
As shown in the zoom-in figures [Fig.~\ref{fig2_band}(b) and~\ref{fig2_band}(e)], the upward dispersive band crosses with the upper flat band but avoids crossing with the lower flat band. The band crossings are unavoidable in both materials, because the two crossed bands belong to different representations of the crystal symmetry group. Take Zn$_2$In$_2$S$_5$($R3m$) as an example, the two bands belong to 2D $\Lambda_4$ and 1D $\Lambda_5$ /$\Lambda_6$ representations, respectively, as distinguished by $C_{3v}$ rotational symmetry around the $k_z$ axis. The different representation prohibits hybridization between them, resulting in a pair of 3D Dirac points at $\pm(0.254, 0.254, 0.254)$ (in units of $2\pi/a$). 
For Zn$_2$In$_2$S$_5$($P6_3mc$), the upper flat band and the upward dispersive band belong to 2D irreducible representation $\Delta_9$ and $\Delta_7$, respectively, of the $C_{6v}$ symmetry. One unique feature of the $P6_3mc$ structure is that there are actually adjacent double Dirac points $(0, 0, \pm 0.306)$ and $(0, 0, \pm 0.308)$ [in units of $(2\pi/a,2\pi/a,2\pi/c)$] with an energy difference of $1.0$ meV. 
This is because there are actually two flat $\Delta_9$ bands that are close to each other but not exactly degenerate [see the inset of Fig.~\ref{fig2_band}(e)].
Since neither $R3m$ nor $P6_3mc$ structure has inversion symmetry, these Dirac semimetal states have fourfold degenerate Dirac points, but with splitting of in-plane band dispersions away from Dirac points [Fig.~\ref{fig2_band}(c) and \ref{fig2_band}(f)]. This is an unique feature of Zn$_2$In$_2$S$_5$, which is different from other Dirac semimetals that require both time-reversal and inversion symmetries. Additionally, we also investigate the strain effect on the electronic structure and found that the position of type-III Dirac points in the $k_z$ axis can be effectively tuned by external strain.

To further reveal the nature of the type-III Dirac points, we fit the energy spectrum from first-principles calculations to a low-energy effective model \footnotemark[\value{footnote}].
If neglecting the tiny splitting induced by inversion symmetry breaking which is insignificant to the main conclusion, the quasiparticles are described by a pair of Weyl Hamiltonian in the vicinity of one Dirac point,
\begin{equation}
h_{\pm}^c=c_\perp(k_x\sigma_x \pm k_y\sigma_y)+c_\parallel \delta k_z\sigma_z+v\delta k_z\sigma_0,
\label{eqHam}
\end{equation}
where $\delta k_z\!=\!k_z-k_z^c$, with $k_z^c\!=\!0.102$ \AA$^{-1}$. The parameters $c_\perp\!=\!2.29$ and $v\!=\!-c_\parallel\!=\!1.36$ eV\AA, indicate that there exists a flat band along $\hat{k}_z$ direction. It is straightforward to derive the topological invariant using $K\!=\!\pm\sigma_z$ in Eq.~(\ref{N2}) and find that $N_2\!=\! 1$. We thus conclude that Zn$_2$In$_2$S$_5$ is a type-III Dirac semimetal.

\begin{figure}
\includegraphics[width =0.8\columnwidth]{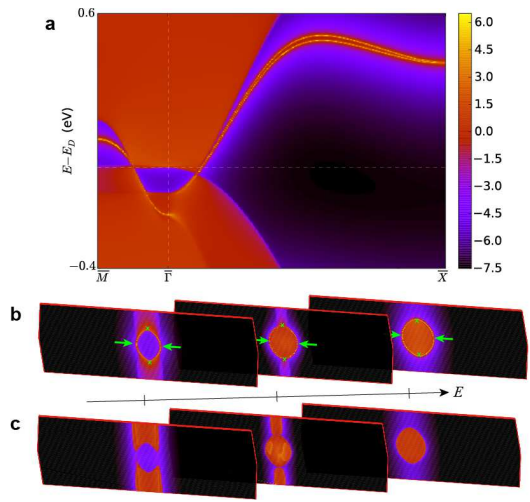}
\caption{\label{fig3_surf} (a) The projected surface density of states for the (100) surface of Zn$_2$In$_2$S$_5$($R3m$) where the nontrivial topological surface states originating from the surface projection of bulk Dirac point are clearly visible. (b) and (c) Constant energy contours of the (100) surface and the bulk at $E=E_F-25$, $E_F$ and $E_F+25$ meV, respectively. The green arrows and crosses mark two pieces of Fermi arcs and the surface projection of bulk Dirac points, respectively.
}
\end{figure}

The topological nature of the type-III Dirac point in Zn$_2$In$_2$S$_5$ is also confirmed by calculating the $\mathbb{Z}_2$ topological invariants which are well-defined in both the $k_z\!=\!0$ and the $k_z\!=\!\pi$ planes. Taking Zn$_2$In$_2$S$_5$($R3m$) as an example, the $k_z\!=\!0$ plane is found topologically nontrivial with $\mathbb{Z}_2\!=\!1$; while the $k_z\!=\!\pi$ plane is topologically trivial with $\mathbb{Z}_2\!=\!0$. Therefore, a band ordering inversion between $\Gamma_4$ and $\Gamma_5+\Gamma_6$ bands should occur along the $k_z$ direction [see Fig.~\ref{fig2_band}(b)], resulting in a band gap closure at the Dirac point.

Topological surface states and Fermi arcs are expected to appear on side surfaces of Zn$_2$In$_2$S$_5$. Figure~\ref{fig3_surf} shows the projected surface DOS for the (100) surface of a semi-infinite Zn$_2$In$_2$S$_5$($R3m$) system, from which we have an intuitive visualization of topological surface state and Fermi arcs. It is seen that the topological surface state emanates from one projection of bulk Dirac point on the (100) surface,
as shown in Fig.~\ref{fig3_surf}(a). The Fermi surface contains two pieces of half-circle Fermi arcs, as shown in Fig.~\ref{fig3_surf}(b), touching at two singularity points where the surface projections of bulk Dirac points appear. Due to the flat Dirac-line Fermi surface, the shape of electron and hole pockets varies rapidly with the increasing chemical potential. All these characteristics, in sharp contrast to conventional metals and topological insulators, should be experimentally observable by modern angle-resolved photoemission spectroscopy technique.

We have shown the existence of type-III Dirac semimetal state in Zn$_2$In$_2$S$_5$. Now we discuss the possibility of realizing a solid-state analogue of black-hole horizon in Zn$_2$In$_2$S$_5$. So far, various black-hole analogues based on different systems have been proposed \cite{novello2002artificial}. For example, Unruh proposed a sonic black hole for sound wave propagating in flowing liquid \cite{PhysRevLett.46.1351, *PhysRevD.51.2827, *visser1998acoustic}, while Volovik suggested a black-hole/white-hole pair in superfluid Helium with a moving vierbein domain well \cite{PhysRevD.58.064021,*volovik2001superfluid, volovik2003universe}. In the theoretical model Eq.~(\ref{eqHam}), the last term is the same as the Doppler shift for quasiparticles under a Galiean transformation to a moving frame of reference with a velocity $v$, which is similar to the case of sonic black hole analogue where the sound wave propagates in moving fluid. In general relativity, a relativistic quasiparticle in 3+1 dimensional spacetime can be described by the line element $ds^2=g_{\mu\nu}dx^\mu dx^\nu$, where $g_{\mu\nu}$ is the inverse (covariant) metric describing an effective curved spacetime in which the relativistic quasiparticles propagate \cite{gron2007einstein}. To obtain a spacetime interpretation, we derive the effective covariant metric $g_{\mu\nu}$ according to Eq.~(\ref{eqHam}) \footnotemark[\value{footnote}]:
\begin{equation}
g_{\mu\nu}=\left(
\begin{array}{cccc}
-(1-v^2/c_\parallel ^2) & 0 & 0 & -v/c_\parallel^2 \\
0 & 1/c_\perp^2 & 0 & 0 \\
0 & 0 & 1/c_\perp^2 & 0 \\
-v/c_\parallel^2 & 0 & 0 & 1/c_\parallel^2
\end{array}
\right),
\end{equation}
which has a similar form of the acoustic metric of Unruh's sonic black hole \cite{PhysRevLett.46.1351, *PhysRevD.51.2827, *visser1998acoustic, novello2002artificial}.

\begin{figure}
\includegraphics[width =0.9\columnwidth]{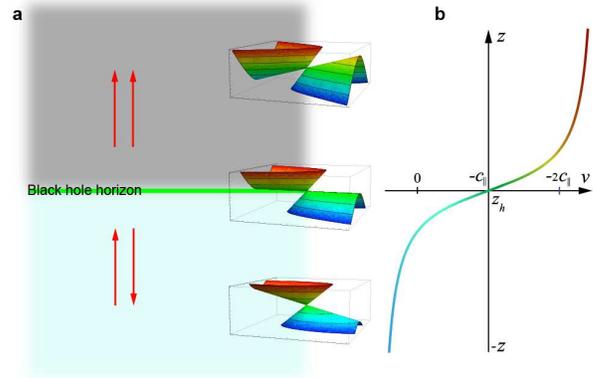}%
\caption{\label{fig4_horizon} (a) Schematic illustration of the solid-state analogue of black hole horizon in an inhomogeneous Zn$_2$In$_2$S$_5$ with controllable structural distortion. The red arrows indicate quasiparticle propagating directions in each region. (b) The dependence of the dragging velocity $v$ on the $z$ coordinate.
}
\end{figure}

Now let's assume that the dragging velocity $v=v(z)$ depends on the spacial $z$ coordinate in an inhomogeneous Zn$_2$In$_2$S$_5$ system, which can, in principle, be realized by controllable (tunable) structural distortion. As the metric has translation invariance in the $x$- and $y$-direction, for simplicity we make a dimension reduction to the 1+1 dimensional spacetime by ignoring the coordinates $x$ and $y$. As a result, the corresponding linear element becomes
\begin{equation}
ds^2=-\left(1-\frac{v^2(z)}{c_\parallel^2}\right)d\tau^2+\frac{dz^2}{c_\parallel^2-v^2(z)}.
\end{equation}
By performing a coordinate transformation: $\tau=t+\int^z dz v(z)/(c_\parallel^2-v^2(z))$, we obtain an effective line element that shares the same form of the radial part of the Schwarzschild line element for gravitational black holes \cite{gron2007einstein}. Similar to the Schwarzschild metric which has a singularity at the Schwarzschild radius corresponding to an event horizon, the above matric also has a horizon ($z_{h}$), where the dragging velocity equals to the local ``speed of light" for quasiparticles: $v(z_{h})=\pm c_\parallel$. The corresponding ``Newtonian gravitational field'' at horizons is given by: $E_g(z_{h}) = \frac{v(z_h)}{c_\parallel^2}\left.\frac{dv}{dz}\right|_{z_{h}}$. According to the fitted parameters of Eq.~(\ref{eqHam}), assuming $v(z)>-c_\parallel$ ($v(z)<-c_\parallel$) in the region $z>z_h$ ($z<z_h$), an inhomogeneous Zn$_2$In$_2$S$_5$ system can be derived (see Fig.~\ref{fig4_horizon}). Hence all quasiparticles in the upper region ($z>z_{h}$) move upward, and cannot cross the plane $z\!=\!z_{h}$, which indicates that this plane is the black-hole horizon. Consequently, the inner observers living in the lower region ($z<z_h$) cannot obtain any information from the upper region ($z>z_{h}$) if they can only use the relativistic quasiparticles for communication.

A black hole can slowly radiate away its mass by emitting a thermal flux at the horizon, as pointed out by Hawking \cite{hawking1974black,*hawking1975particle}. The analogous model presented above not only suggests a new route to simulating an event horizon, but also facilitates the realization of the Hawking radiation analogue in inhomogeneous type-III Dirac semimetals. Although this model is static in equilibrium, the dissipation process right after the creation of the black hole horizon analogue is similar to the process of the Hawking radiation \cite{volovik2016black, volovik2016lifshitz, volovik2017exotic}. The corresponding Hawking temperature can be directly determined by the ``surface gravity'' at the horizon $E_g(z_h)$  \cite{volovik2003universe, novello2002artificial}:
\begin{equation}
T_H=\frac{\hbar c_\parallel}{2\pi k_B}E_{g}(z_{h})=\frac{\hbar}{2\pi k_B}\left.\frac{dv}{dz}\right|_{z_{h}},
\end{equation}
where $k_B$ and $\hbar$ are the Boltzmann and the reduced Planck constants, respectively. Obviously, $T_H$ may reach high temperature as long as the gradient of the dragging velocity is sufficiently large across the horizon. As the dragging velocity $v$ is a material-dependent parameter which can be effectively tuned by strain and chemical doping \cite{le2016three,guan2016strain}, it is expected to reach a high value of $T_H$ in inhomogeneous Zn$_2$In$_2$S$_5$.

In addition to simulating event horizons and Black hole evaporation, 
other novel astrophysical phenomena such as gravitational lensing effect \cite{guan2016strain}, gravity wave \cite{PhysRevD.66.044019}, and cosmological constant problem \cite{volovik2001superfluid} can also be explored in type-III Dirac semimetals.
In future work we will investigate possible strain engineering of the black-hole horizon analogue in inhomogeneous In$_2$Zn$_2$S$_5$. Other interesting directions for future research including the realization of other topological states, such as type-III Weyl semimetal by breaking time-reversal symmetry via magnetic doping, the interplay between different topological states in Zn$_2$In$_2$S$_5$, and the analytical estimation of transport and optical behaviors of type-III topological semimetals.

\begin{acknowledgments}
This work was supported by DOE-BES (Grant No. DE-FG02-04ER46148). The calculations were done on the CHPC at the University of Utah and DOE-NERSC.
\end{acknowledgments}

\providecommand{\noopsort}[1]{}\providecommand{\singleletter}[1]{#1}%

\end{document}